\documentclass[preprint, showkeys, showpacs, preprintnumbers, amsmath, amssymb, floatfix]{revtex4}
\usepackage{graphicx}
\usepackage{dcolumn}
\usepackage{bm}

\begin{document}

\title{Thermoelectric power and Hall coefficient measurements on Ba(Fe$_{1-x}$$TM$$_x$)$_2$As$_2$ ($TM$ = Co and Cu)}

\author{Eun Deok Mun, Sergey L. Bud'ko, Ni Ni, and Paul C. Canfield}
\affiliation{Ames Laboratory U.S. DOE and Department of Physics and Astronomy, Iowa State University, Ames, Iowa 50011, USA}

\date{\today}

\begin{abstract}

Temperature dependent thermoelectric power (TEP) data on Ba(Fe$_{1-x}$$TM$$_x$)$_2$As$_2$ ($TM$ = Co and Cu), complemented by the Hall coefficient data on the samples from the same batches, have been measured.  For Co-doping we clearly see a change in the temperature dependent TEP and Hall coefficient data when the sample is doped to sufficient $e$ (the number of extra electrons associated with the $TM$ doping) so as to stabilize low temperature superconductivity.  Remarkably, a similar change is found in the Cu-doped samples at comparable $e$-value, even though these compounds do not superconduct.
These changes possibly point to a significant modification of the Fermi surface / band structure of  Ba(Fe$_{1-x}$$TM$$_x$)$_2$As$_2$ at small electron doping, that in the case of Co-doping is just before, and probably allows for, the onset of superconductivity.
These data further suggest that suppression of the structural / magnetic phase transition and the establishment of a proper $e$-value are each necessary but, individually, not sufficient conditions for superconductivity.

\end{abstract}


\maketitle

\section{Introduction}

Superconductors with high transition temperature ($T_c$) values have, for decades, attract attention from different parts of the condensed matter and applied physics communities. Recently this interest was re-ignited by the discovery of superconductivity with $T_c$ up to $\sim 55$ K in new, Fe-As based materials. \cite{kam08a,che08a,ren08a,rot08a} Two families of such materials, $R$FeAsO ($R$ = rare earth) and $AE$Fe$_2$As$_2$ ($AE$ = Ba, Sr, Ca) are currently being explored in a great detail. In both cases superconductivity can be induced (or enhanced) by either doping (both by electrons and holes) or application of pressure. Despite the advantage of achieving higher $T_c$ values, the synthesis and doping in the $R$FeAsO family so far appears to be complex and on many counts difficult to control and reproduce. On the other hand, the $AE$Fe$_2$As$_2$ family, and the electron-doped Ba(Fe$_{1-x}$$TM$$_{x}$)$_{2}$As$_{2}$ ($TM$ = Co, Ni, Cu, Rh, Pd) series in particular, can be synthesized in a single crystal form, are highly reproducible, and offer maximum in $T_c$ values in excess of 20 K as well as a region of coexistence of magnetism and superconductivity. \cite{sef08a,nin08a,chu09a,nin09a,lil09a,can09a,nin09b} These series have recently become an archetypical set of materials for studies of Fe-As superconductivity.

The phase diagrams of the Ba(Fe$_{1-x}$$TM$$_{x}$)$_{2}$As$_{2}$ series are well established for $3d$- and $4d$-$TM$. \cite{can09a,nin09b} In the pure BaFe$_2$As$_2$ the first order, structural and antiferromagnetic transitions coincide in temperature (at $\sim 135$ K). With low Co or Cu doping these phase transitions are suppressed and split into two distinct transitions, the higher temperature one being structural and the lower temperature one being magnetic. \cite{pra09a,les09a} For Co-doping, superconductivity was observed in the range of concentrations between $x \sim 0.035$ and $x \sim 0.17$. For $0.035 \leq x \leq 0.06$ magnetism and superconductivity coexist.  For Cu-doping, on the other hand, superconductivity was not observed, even when the structural and magnetic phase transitions were suppressed to values similar to or even lower than the Co-doped analogs that do superconduct.  Whereas the upper, structural / antiferromagnetic, phase transitions are suppressed in a similar manner by $x$, the amount of $TM$ dopant, the superconducting dome appears to be limited to a specific region of $e$-values, where $e$ is the number of extra electrons provided by the $TM$ substitution. \cite{can09a,nin09b} The $T - e$ and $T - x$ phase diagrams shown in Fig. \ref{F1} (i) delineate the region of $e$-values that supports superconductivity and (ii) illustrate the observation that the suppression of the upper, structural / antiferromagnetic, phase transitions is a necessary, but not sufficient, condition to stabilizing superconductivity in these materials:  if the upper transitions are suppressed too slowly, then the window of $e$-values that support superconductivity can be overshot.

In this work, in an effort to better understand the changes induced by $TM$ substitutions, we present temperature dependent thermoelectric power (TEP) studies for different levels of Co- and Cu-doping complemented by the Hall coefficient data on the samples from the same batches.  For Co-doping we clearly see a change in the temperature dependent TEP and Hall coefficient data when the sample is doped to sufficient $e$ so as to stabilize low temperature superconductivity.  Remarkably, a similar change is found in the Cu-doped samples at comparable $e$-value, even though these compounds do not superconduct.

\section{Experimental}

Single crystals of Ba(Fe$_{1-x}$$TM$$_{x}$)$_{2}$As$_{2}$, $TM$ = Co and Cu, were grown out of excess FeAs-flux. \cite{nin08a,can09a} The actual Co and Cu concentration in the crystals was determined by employing wavelength dispersive x-ray spectroscopy. For the transport measurements, the crystals were cut and / or cleaved with a razor blade into dimensions of typically $\sim 0.8 \times 0.07 \times 3$ mm$^{3}$ for TEP and into dimensions of typically $\sim 2.5 \times 0.07 \times 3$ mm$^{3}$ for Hall measurements. Hall resistivity data was collected using the ac transport option of a Quantum Design (QD) Physical Property Measurement System (PPMS) in a four wire geometry with switching the polarity of the magnetic field ($H \| c$) to remove any magnetoresistive components due to the misalignment of the voltage contacts. The current contacts were carefully painted using Epotek H20E silver epoxy to attach Pt wires to cover two opposing side faces of the plate-shaped crystals to ensure as uniform of a current flow as possible. The voltage contacts were placed across from each other on the two remaining side faces of the crystals. Magnetic field dependence of the Hall resistivity is essentially linear over the whole temperature range (see inset of Fig. \ref{F7} below for a typical set of data), so the data taken in 90 kOe applied field represent the temperature dependence of the Hall coefficient fairly well.

Thermoelectric power measurements were carried out by a DC, alternating heating (two-heater and two-thermometer), technique in the temperature
range from 2 K to 300 K using a homemade set-up in a QD PPMS. \cite{mun09a} The samples were directly attached to the two
Cernox thermometers using DuPont 4929N silver paint. The voltage difference, $\Delta V$, between the hot and the cold ends of the sample was measured by a HP 34420A nanovoltmeter.  The voltage leads were phosphor-bronze wire. The temperature difference $\Delta T$ of $\sim 0.3$ K to $\sim 0.75$ K was established using two strain gauge heaters glued next to the sample on the thermometers. The TEP value of phosphor-bronze is ignored since $S$ of this wire is less than 0.5$\mu$V/K for the whole temperature measured. 

It should be noted that the $AE$Fe$_2$As$_2$ ($AE$ = Ba, Sr, Ca) materials are prone to exfoliation along the $c$-axis that can lead to larger than conventionally accepted errors in resistivity measurements due to poorly-defined current path lengths and samples cross-sections. \cite{nin08b,tan09a} In an effort to minimize this we tried to use very thin, un-deformed, cleaved pieces for the Hall measurements.
In contrast, no knowledge of the geometric dimensions of the sample is needed to calculate the thermoelectric power, the only requirement being to measure the temperature gradient and the voltage between the same points of the sample. Additionally, for pure BaFe$_2$As$_2$ and the samples with lower Co and Cu concentrations, possible differences in the structural/antiferromagnetic domain \cite{tan09b} distribution in the ordered state may cause some (small) differences in the measured Hall coefficient and TEP at low temperatures. This being said, we do not see any evidence for this being a large, or poorly controlled, effect.

\section{Data and Results}

\begin{figure}
\begin{center}
\includegraphics[angle=0,width=90mm]{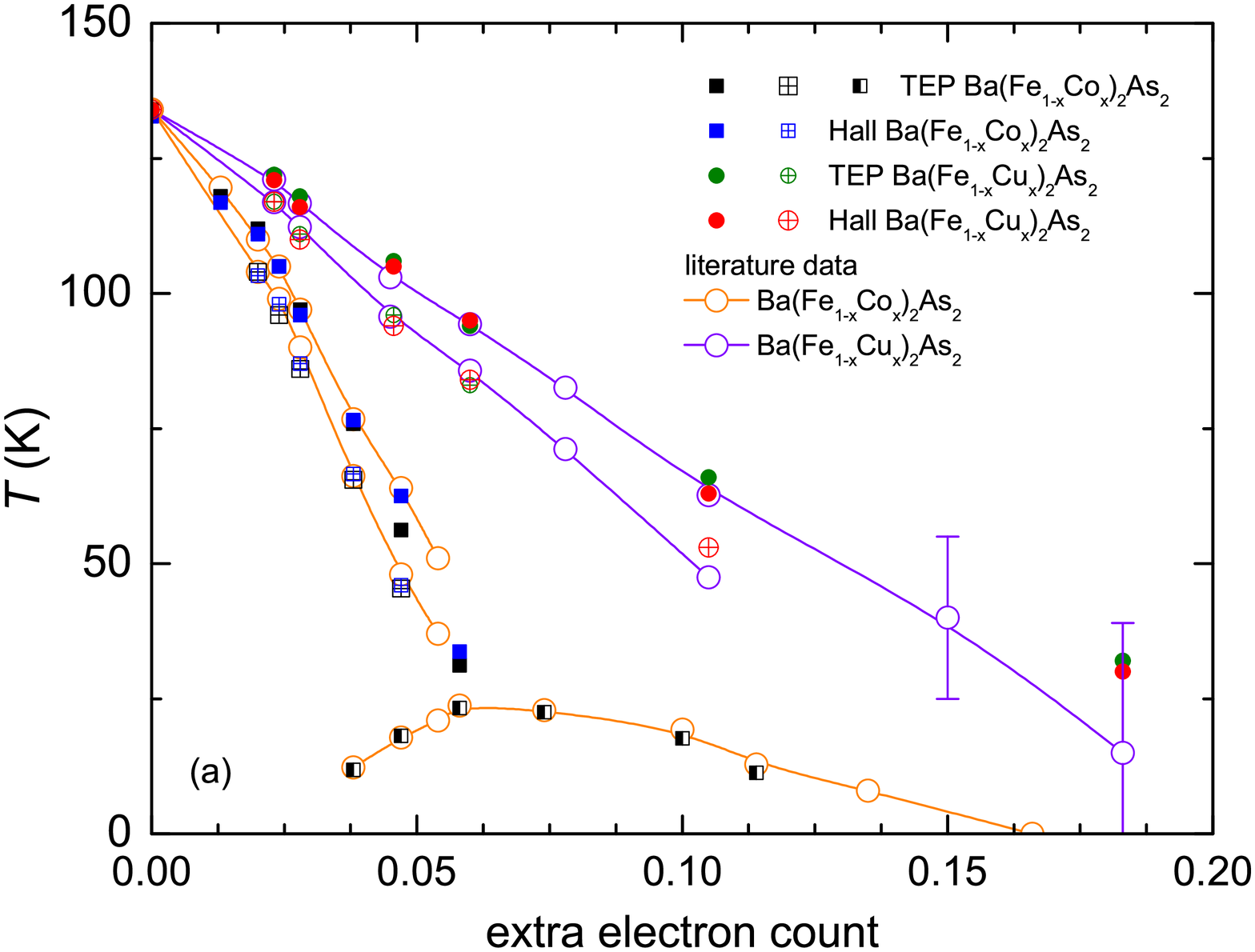}
\includegraphics[angle=0,width=90mm]{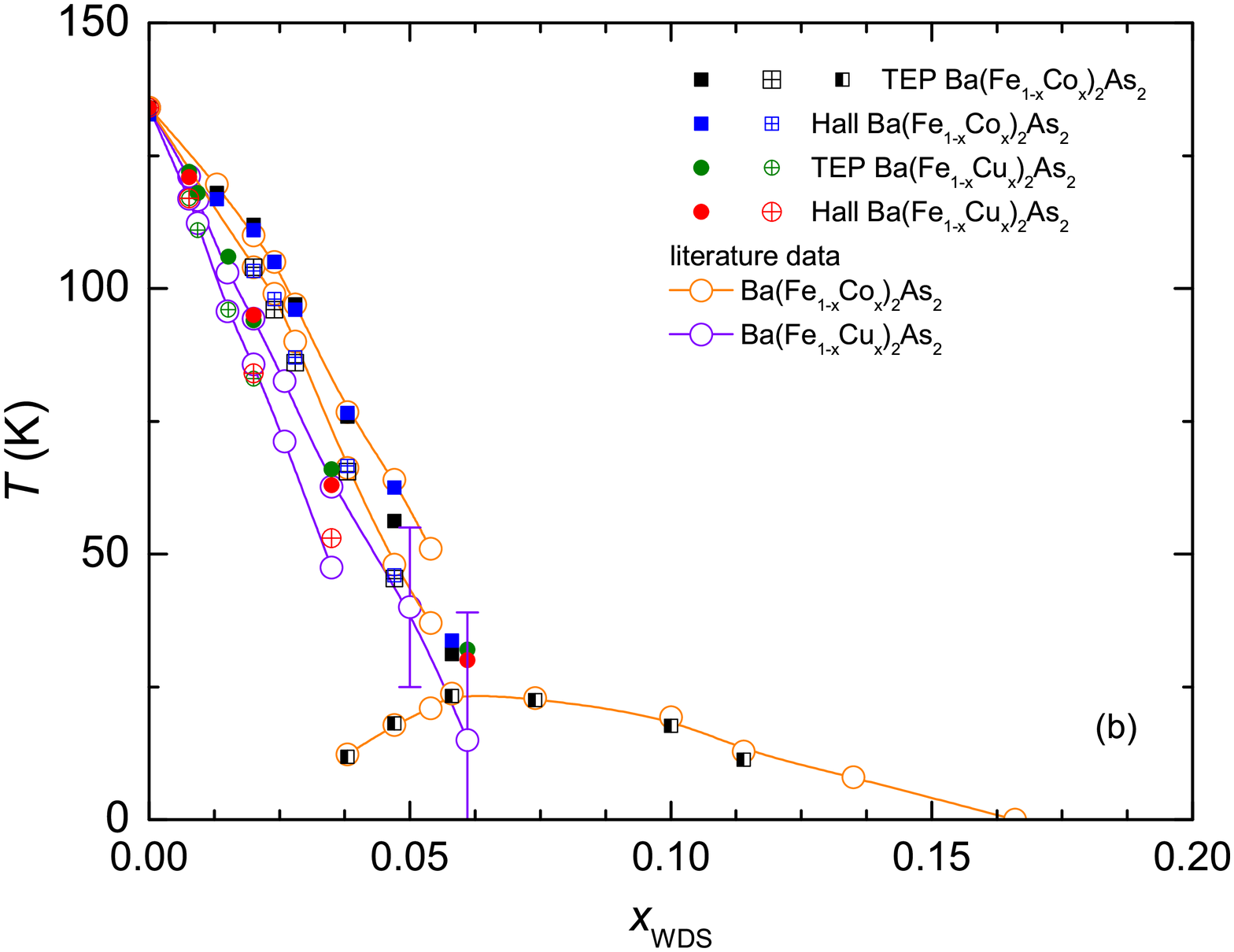}
\end{center}
\caption{(Color online) Transition temperature as a function of (a) extra electrons per Fe / $TM$ site, and (b) measured, $x_{WDS}$, $TM$ concentration, phase diagrams for Ba(Fe$_{1-x}$Co$_x$)$_2$As$_2$ and Ba(Fe$_{1-x}$Cu$_x$)$_2$As$_2$ from references \onlinecite{nin08a,can09a}.  Data points (filled - structural transition, crossed - magnetic transition, and half-filled - superconducting transition) inferred from TEP and Hall coefficient data (shown below) are also shown and often overlap literature data.}\label{F1}
\end{figure}

The temperature dependent TEP and Hall coefficient data for Ba(Fe$_{1-x}$Co$_x$)$_2$As$_2$ are shown for $x \leq 0.114$ in Figs. \ref{F3} and \ref{F2}.  These data clearly show the suppression of the upper, structural/magnetic, phase transition as well as the lower temperature superconductivity and further support the $T - x$ (and $T - e$) phase diagrams presented in Refs. [\onlinecite{nin08a,chu09a,nin09a,can09a}] and shown in Fig. \ref{F1}.  The Hall data (Fig. \ref{F3}) show a clear break in slope at the higher temperature, structural / magnetic transition that is systematically suppressed as $x$ increases until it is no longer easily detectable for $x > 0.047$.  For $x \geq 0.38$ superconductivity manifests itself, but given the 90 kOe applied field, $T_c$ is slightly suppressed (consistent with $H_{c2}$ data) from the values found in Ref. [\onlinecite{nin08a}] and shown in Fig. \ref{F1}.  The TEP data (Fig. \ref{F2}) also show an anomaly that is systematically suppressed with increasing $x$, but for $x \geq 0.038$ it is increasingly subtle and can only be clearly seen in $dS/dT$ plots.  On the other hand, since TEP data can be collected in zero applied magnetic field, $T_c$ is clearly seen and in excellent agreement with the points shown in Fig. \ref{F1}.

As $x$ is increased the signatures of the structural and magnetic phase transitions become less pronounced, especially in the TEP data.  In order to consistently extract transition temperatures we adopt similar, derivative criteria as were used for resistivity data \cite{nin08a} and subsequently supported by diffraction measurements \cite{pra09a}.  Figure \ref{F4} presents these criteria for $x = 0.047$.  The data points inferred in this manner are shown in Fig. \ref{F1} and agree well with the data inferred from resistivity, susceptibility and specific heat measurements. \cite{nin08a,can09a}

The more conspicuous aspect of Figs. \ref{F3} and \ref{F2}, though, is the clear difference between the $x \leq  0.020$ data and the $x \geq 0.038$ data.  This is seen most strikingly in the TEP data where the $x \leq 0.020$ and $x \geq 0.024$ data sets appear to fall onto two separate manifolds over the whole measured temperature range.  For the Hall data there again is a distinct difference between the $x \leq 0.024$ and the $x \geq 0.038$ data sets:  for $x \leq 0.024$  there seems to be a similar, low temperature value of approximately - 0.3 n$\Omega$ cm/Oe whereas for the $x \geq 0.038$ the low temperature value, before the onset of superconductivity drives the data to zero, rises and finally saturates near approximately - 0.02 n$\Omega$ cm/Oe. The $x = 0.028$ data lay in between. This change in the low temperature Hall data can be seen more quantitatively in Fig. \ref{F5} where the Hall coefficient just above the maximum $T_c$ value ($T = 25$ K) is is plotted as a function of $e$.  It is noteworthy that the Hall coefficient data presented in Fig. \ref{F3} is quantitatively similar to the data presented in Ref. \onlinecite{fan09a} as well as in Ref. \onlinecite{rul09a} for the limited subsets of samples with similar $x$ values. 

In the case of Co-doping the change in the overall form of the TEP and Hall data takes place as the sample is doped into the region of $e$-values that support superconductivity.  In order to determine whether this behavior is intractably linked to superconductivity, or is a more generic feature of electron doping in Ba(Fe$_{1-x}$$TM$$_x$)$_2$As$_2$ compounds we performed similar measurements on Ba(Fe$_{1-x}$Cu$_x$)$_2$As$_2$.

The temperature dependent TEP and Hall coefficient data for Ba(Fe$_{1-x}$Cu$_x$)$_2$As$_2$ are shown for $x \leq 0.061$ in Figs. \ref{F7} and \ref{F6}.  The Hall data (Fig. \ref{F7}) show a clear break in slope at the higher temperature structural / magnetic transition that is systematically suppressed as $x$ increases until by $x = 0.061$ when the break only manifests itself weakly.  The TEP data (Fig. \ref{F6}) also show an anomaly that is systematically suppressed with increasing $x$, but for $x \geq 0.020$ it is increasingly subtle.  The suppression of the structural / antiferromagnetic phase transitions as well as the fact that neither the Hall nor the TEP data show any signature of superconductivity are consistent with the results of Ref. \onlinecite{can09a} and, as shown in Fig. \ref{F1}, the structural / antiferromagnetic transition temperatures inferred from the TEP and Hall coefficient measurements are in good agreement with those inferred from resistivity measurements. \cite{can09a}

Remarkably, Ba(Fe$_{1-x}$Cu$_x$)$_2$As$_2$ manifests similarly dramatic changes in behavior as Cu-doping is increased, as were found for Ba(Fe$_{1-x}$Co$_x$)$_2$As$_2$, even though Ba(Fe$_{1-x}$Cu$_x$)$_2$As$_2$ does not have any onset of superconductivity associated with this change.
The TEP data again have two regimes:  $x \leq 0.0077$ and $x \geq 0.0093$ with the data falling onto one of two separate manifolds over the whole measured temperature range.  For the Hall data there again is a distinct difference between the $x \leq 0.0093$ and the $x \geq 0.015$ data sets:  for $x = 0$ and 0.0077 the low temperature Hall coefficient values are close to approximately - 0.3 n$\Omega$ cm/Oe, for $x = 0.0093$ this value is close to -0.25 n$\Omega$ cm/Oe, whereas for the $x \geq 0.020$ the low temperature value rises and finally saturates near  approximately - 0.05 n$\Omega$ cm/Oe.  The change in the low temperature Hall data for both series can be seen more quantitatively in Fig. \ref{F5}.  Given that there is no superconductivity in the measured Ba(Fe$_{1-x}$Cu$_x$)$_2$As$_2$ samples, we can see that using the $T = 25$ K value of $\rho_H/H$ is a valid approximation for the zero temperature extrapolation and allows coparison with the Co-doped data.

\section{Analysis and Discussion}

\begin{figure}
\begin{center}
\includegraphics[angle=0,width=120mm]{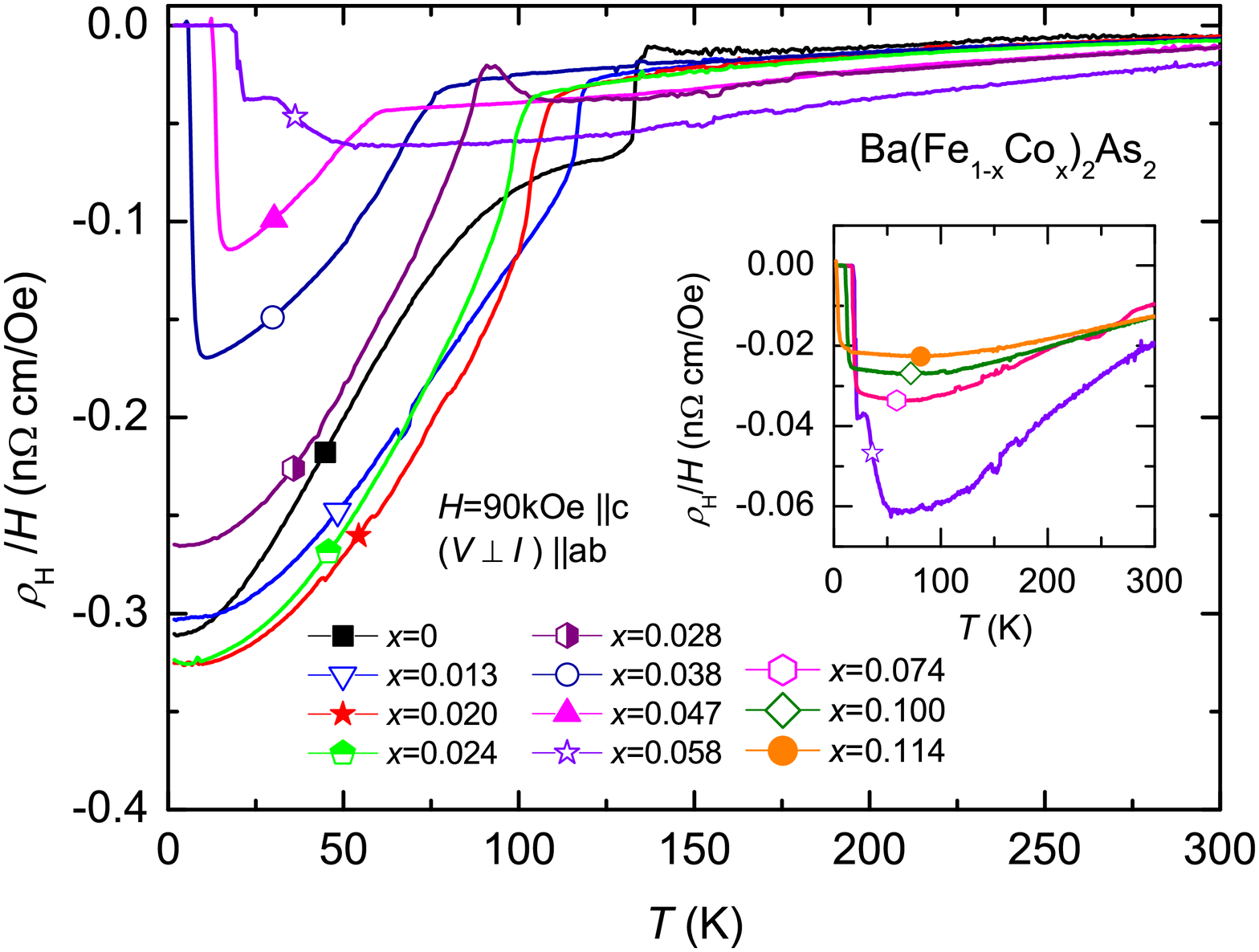}
\end{center}
\caption{(Color online) $\rho_H/H$ (Hall coefficient) as a function of temperature for Ba(Fe$_{1-x}$Co$_x$)$_2$As$_2$.  Inset: enlarged scale to show data for higher $x$ values.}\label{F3}
\end{figure}

\begin{figure}
\begin{center}
\includegraphics[angle=0,width=120mm]{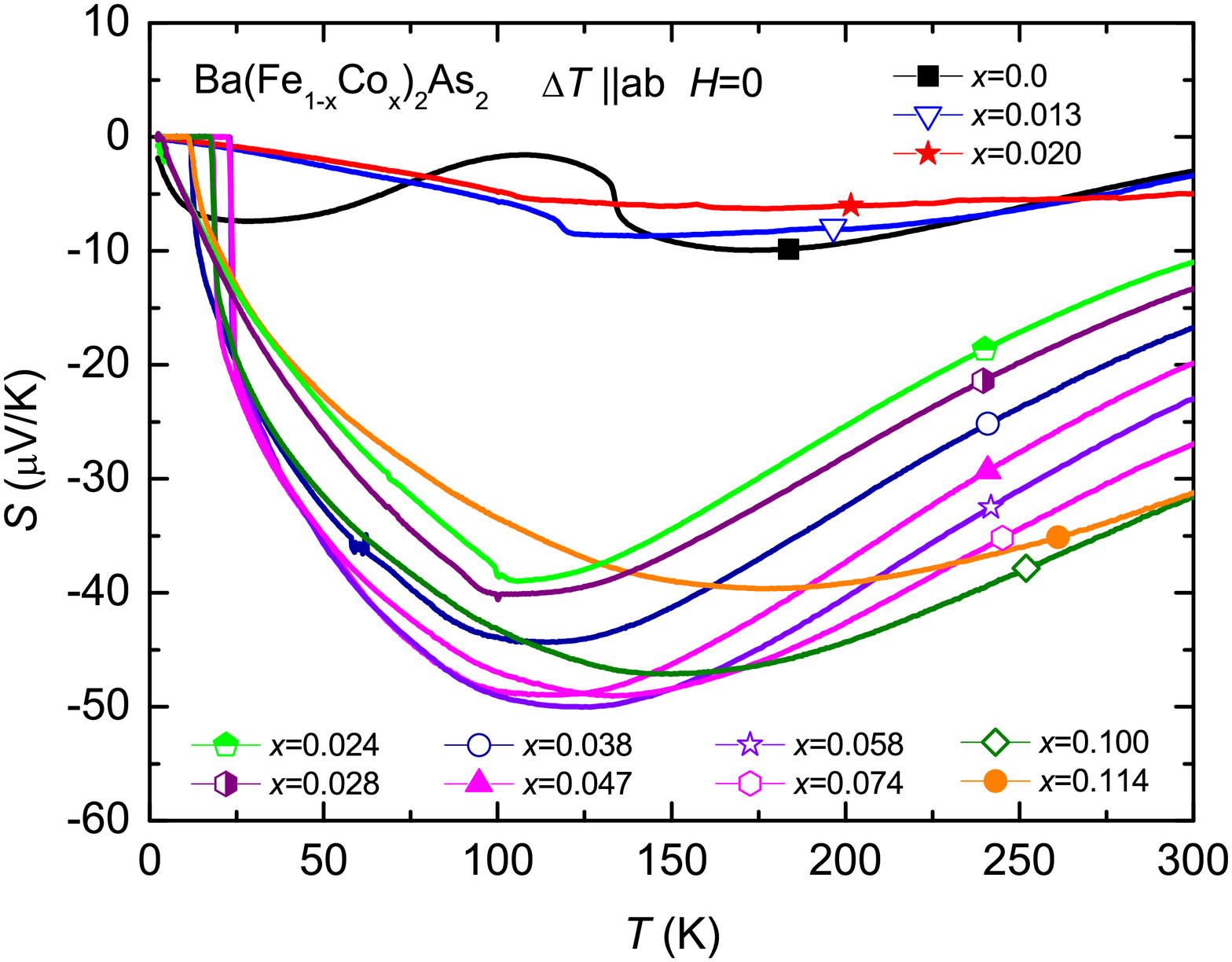}
\end{center}
\caption{(Color online) Thermoelectric power as a function of temperature for Ba(Fe$_{1-x}$Co$_x$)$_2$As$_2$.}\label{F2}
\end{figure}

\begin{figure}
\begin{center}
\includegraphics[angle=0,width=80mm]{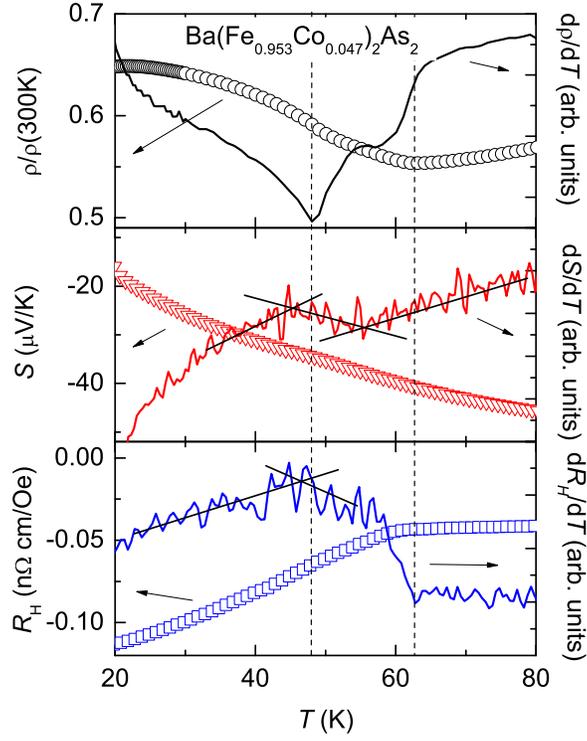}
\end{center}
\caption{(Color online) Derivative criteria used to infer upper (structural) and lower (magnetic) phase transitions from transport data (see Refs. \onlinecite{nin08a,pra09a} for further discussion).  The dotted lines are the values of the transition temperatures inferred from the resistivity data.  It should be noted that the transition inferred from the TEP data for this Co concentration are slightly lower than those inferred from the resistivity and Hall coefficient data.}\label{F4}
\end{figure}

The TEP and Hall coefficient data presented in Figs. \ref{F3}, \ref{F2}, \ref{F7}, and \ref{F6} (i) confirm the established structural / antiferromagnetic and superconducting (or lack there of) phase lines for the Ba(Fe$_{1-x}$Co$_x$)$_2$As$_2$ and Ba(Fe$_{1-x}$Cu$_x$)$_2$As$_2$ series and (ii) indicate that there appears to be distinct change in the electronic properties of these compounds associated with increasing the $e$-value beyond $\sim 0.020$.  Whereas both points can be inferred from either measurement, it is worth noting that the structural / antiferromagnetic phase transitions remain more clearly seen in the Hall coefficient data whereas the distinct change in the electronic properties with increasing $e$ is more clearly seen in the TEP data.

\begin{figure}
\begin{center}
\includegraphics[angle=0,width=120mm]{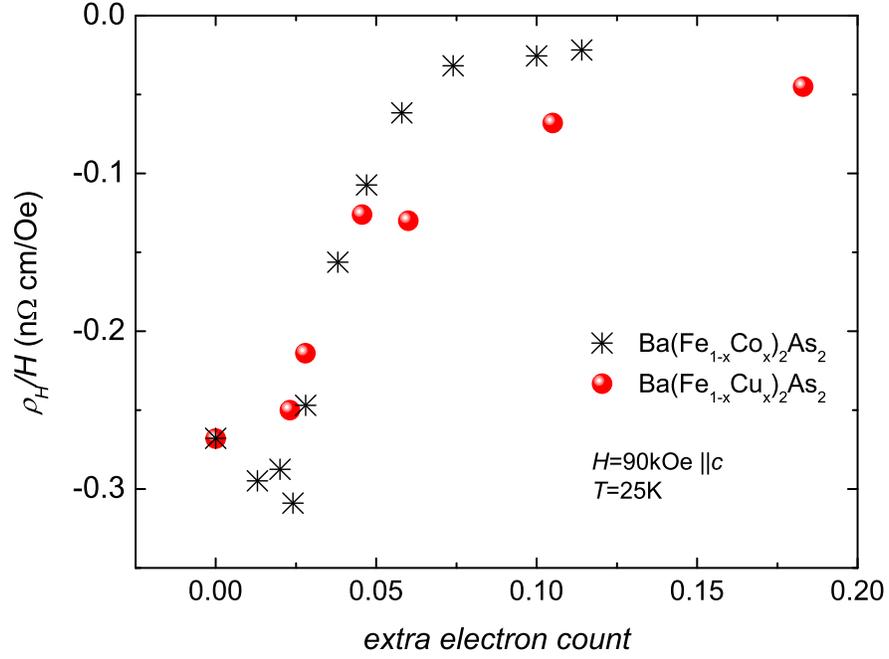}
\end{center}
\caption{(Color online) The low temperature ($T = 25$ K) Hall coefficient data as a function of extra electron count, $e$, for Ba(Fe$_{1-x}$Co$_x$)$_2$As$_2$ and Ba(Fe$_{1-x}$Cu$_x$)$_2$As$_2$.}\label{F5}
\end{figure}

\begin{figure}
\begin{center}
\includegraphics[angle=0,width=120mm]{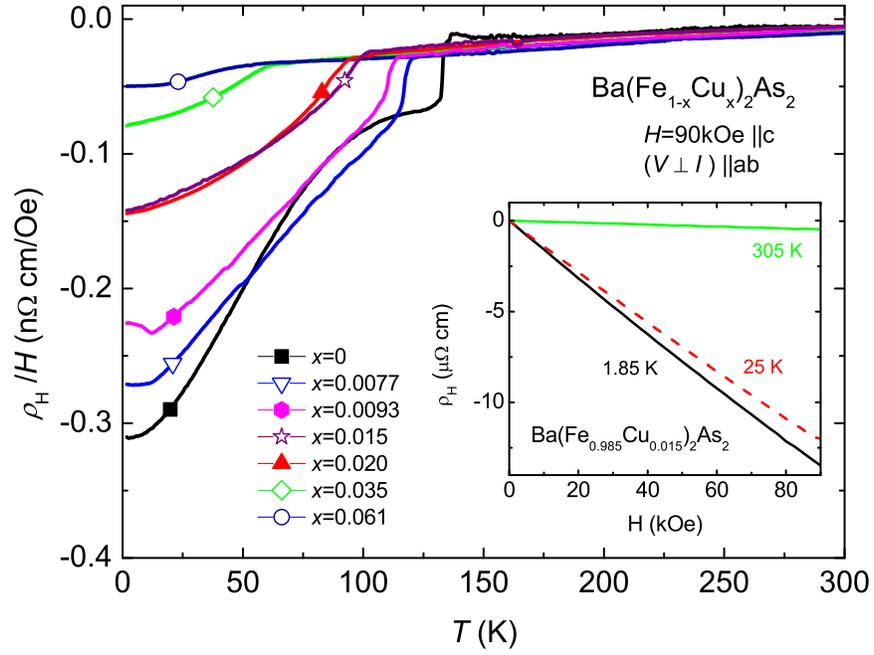}
\end{center}
\caption{(Color online) $\rho_H/H$ (Hall coefficient) as a function of temperature for Ba(Fe$_{1-x}$Cu$_x$)$_2$As$_2$. Inset: field-dependent Hall resistivity, $\rho_H$, of Ba(Fe$_{0.985}$Cu$_{0.015}$)$_2$As$_2$ measured at 1.85 K, 25 K, and 305 K.}\label{F7}
\end{figure}

\begin{figure}
\begin{center}
\includegraphics[angle=0,width=120mm]{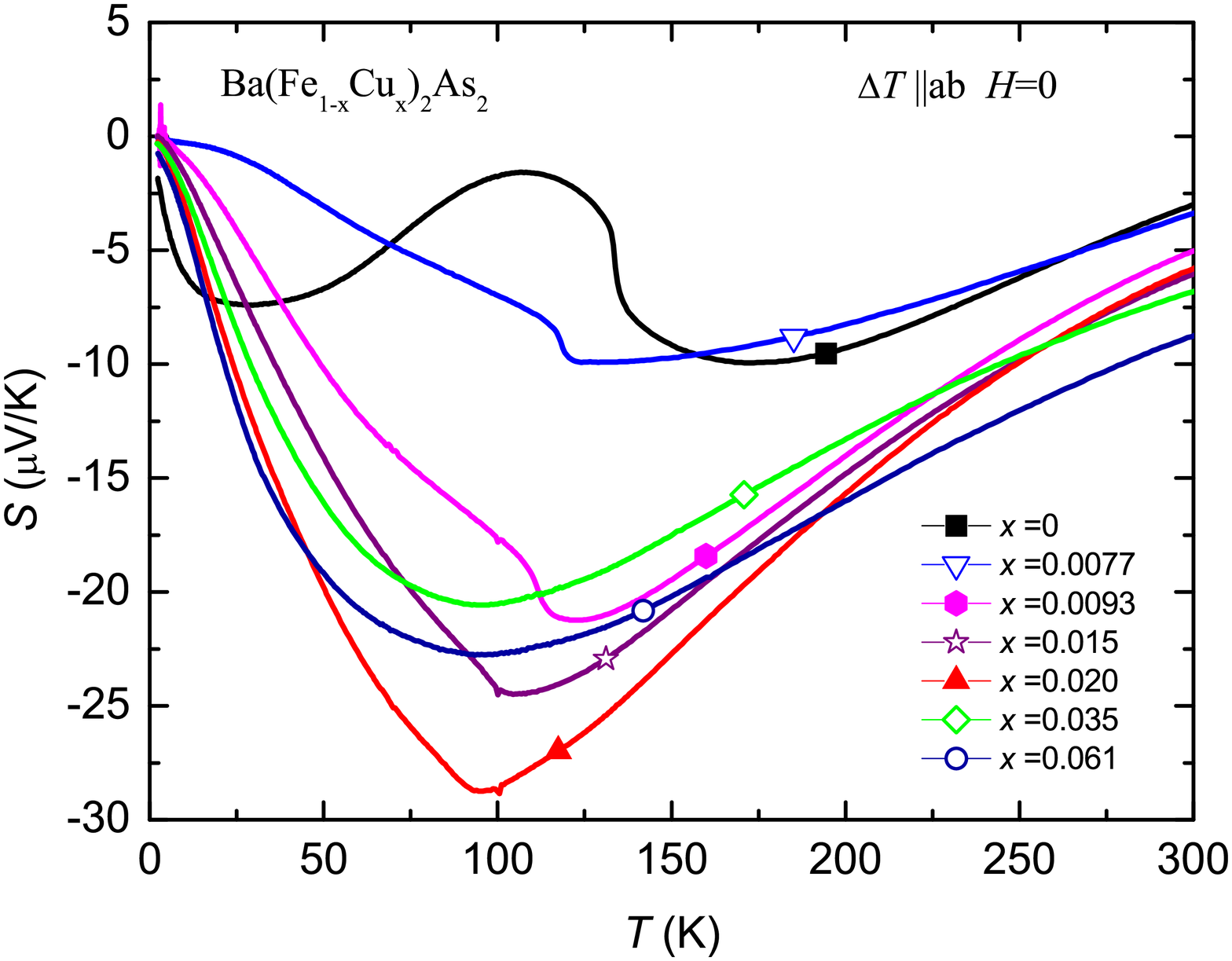}
\end{center}
\caption{(Color online) Thermoelectric power as a function of temperature for Ba(Fe$_{1-x}$Cu$_x$)$_2$As$_2$.}\label{F6}
\end{figure}

Both Ba(Fe$_{1-x}$Co$_x$)$_2$As$_2$ and Ba(Fe$_{1-x}$Cu$_x$)$_2$As$_2$ manifest dramatic changes in TEP (over the whole temperature range) at a certain concentrations of the transition metal dopants (Figs. \ref{F2} and \ref{F6}). It is noteworthy that the ratios of Co doping value to Cu doping value ($x_{Co}/x_{Cu}$) for the highest concentration on a low- doping manifold (0.020/0.0077), lowest concentration on the high-doping manifold (0.024/0.0093) and the average of two (0.022/0.0085) are 2.6, that is very close to what is expected in case of $e$ (extra electron) scaling if the valence of Cu is the same as that of Fe, Co or Ni and the extra $d$ electrons essentially provide a right band shift. \cite{can09a,nin09b} The low temperature (25 K) Hall coefficient data for Ba(Fe$_{1-x}$Co$_x$)$_2$As$_2$ and Ba(Fe$_{1-x}$Cu$_x$)$_2$As$_2$ essentially lay on the same line if plotted as a function of $e$ (Fig. \ref{F5}) Whereas the break in TEP behavior is conspicuous, the evolution of the Hall coefficient is somewhat more gradual and the "critical concentration" is more difficult to infer. This said, there is a clear change in behavior of the Hall coefficient and it scales with $e$. Since TEP, grossly speaking, depends on the derivative of the density of states at the Fermi level, it is possibly sensitive to subtle curvature changes of the Fermi surface as a precursor of the topology changes at slightly higher concentrations that are seen in the Hall effect.

Thermoelectric power and Hall coefficient are known to be very sensitive to the Fermi surface topology. \cite{var89a,liv99a} Broadly speaking, measurements of TEP and Hall coefficient probe convolutions of the Fermi surface / band structural properties as well as scattering, especially in a multiband intermetallic compound.  This being said, the dramatic changes seen in the TEP as well as the Hall data are more likely to be associated with changes in the Fermi surface / band structural properties than scattering.  This argument is supported by the idea that there may be some form of topological change or a significant distortion in the Fermi surfaces of the Ba(Fe$_{1-x}$$TM$$_x$)$_2$As$_2$ compounds at a given, small change in the band filling ($e$ - value).  In addition, such a sudden change, specifically in the TEP is hard to associate with a change in scattering.  At a gross level, drawing on the intuition provided by single band models, the fact that the change in TEP is so much more dramatic implies that there may be a more dramatic change in the energy derivative of the density of states near the Fermi level than in the actual density of states itself, but more detailed analysis and modeling will be needed to clarify the origin of the dramatic changes in these measurements with doping.  It should also be noted, that there are qualitative changes in the resistive anomalies at these critical dopings as well.  Figure 1 of Ref. \onlinecite{can09a} shows that for $x \leq  0.020$ for Co, and $x \leq 0.0077$ for Cu,  resistivity data show a sharper, cusp-like resistive anomaly associated with the structural / magnetic phase transitions whereas for higher $x$ the resistive anomaly is broader, monotonically increasing with decreasing temperature, much more rounded or shoulder-like, and the two transitions are increasingly separated.

Based on earlier phase diagram work, \cite{can09a,nin09b} it has been proposed that whereas for Co-doping, when the structural / antiferromagnetic phase transitions are sufficiently suppressed superconductivity is stabilized over a finite range of $e$ - values, for Cu-doping the upper phase transitions are suppressed more slowly (as a function of $e$) and the finite range of $e$ - values that supports superconductivity is overshot, i.e. by the time the upper transitions are suppressed enough the Cu-doped samples no longer have the correct band filling. \cite{can09a,nin09b}   The fact that the same qualitative changes in the TEP and Hall coefficient data occur in both Co-doped and Cu-doped BaFe$_2$As$_2$, independent of the occurrence of low temperature superconductivity, is further evidence of this idea that there are a set of necessary, but not sufficient conditions that have to be met in order to stabilize superconductivity in the Ba(Fe$_{1-x}$$TM$$_x$)$_2$As$_2$ materials.  Both the TEP and Hall coefficient data suggest a change in the Fermi surface / band structural properties near $e \sim 0.020$.  For Ba(Fe$_{1-x}$Co$_x$)$_2$As$_2$ the upper transitions are suppressed sufficiently and superconductivity occurs, on the other hand, for Ba(Fe$_{1-x}$Cu$_x$)$_2$As$_2$ the upper structural and magnetic transitions are still too high and superconductivity is not detected.

It is important to mention again that the Hall coefficient data for Ba(Fe$_{1-x}$Co$_x$)$_2$As$_2$ presented here is in very good agreement with that presented in Refs. \onlinecite{fan09a,rul09a} for the subsets of overlapping concentrations. Indeed, there was a general sense that, "The band structure of BaFe$_2$As$_2$ appears then very fragile as it is disturbed by a small shift of the chemical potential" \cite{rul09a}, but due to sparse Hall effect data and no TEP results, it was not appreciated that there was such a clear critical $e$-doping level. The key differences between this work and these prior studies are:  a higher density of low-$x$ samples, additional measurements on Ba(Fe$_{1-x}$Cu$_x$)$_2$As$_2$, and, very importantly, TEP data.  In a similar manner, recent measurements \cite{lil09a} of TEP on Ba(Fe$_{1-x}$Ni$_x$)$_2$As$_2$ appear to be fully consistent with our conclusions, but the relatively small number of $x$-values studied prevented the discovery of the sudden, dramatic change in TEP as $e$ is increased past 0.020.  More detailed studies of TEP and Hall coefficient on Ba(Fe$_{1-x}$$TM$$_x$)$_2$As$_2$  ($TM$ = Ni, Rh, Pd) will hopefully refine our understanding of how general this behavior is.

In conclusion, the TEP and Hall coefficient data provide clear evidence for a change in the electronic properties of  Ba(Fe$_{1-x}$Co$_x$)$_2$As$_2$  and Ba(Fe$_{1-x}$Cu$_x$)$_2$As$_2$ at an $e$-value close to that associated with the occurrence of superconductivity in other Ba(Fe$_{1-x}$$TM$$_x$)$_2$As$_2$ series. \cite{can09a,nin09b}  These data further demonstrate that suppression of the structural / magnetic phase transition and the establishment of a proper $e$ - value (band filling) are each necessary but, individually, not sufficient conditions for superconductivity. Whereas this work provides a clear condition of the low-$e$ onset of the region that supports superconductivity, the specific effect of lowering the structural / antiferromagnetic transition temperature sufficiently (i.e. reducing the size of the orthorhombic distortion, reducing the size of the ordered moment and/or changing the excitation spectrum) still needs to be identified. In addition, further work, specifically studying the Fermi surface / band structural properties of these series will be needed to clarify the nature of the change taking place for $e \sim 0.020$ as well as to explain the dramatic changes in the TEP.

\begin{acknowledgments}
We thank Adam Kami\'nski and Makariy A. Tanatar for useful discussions. We acknowledge Florence Rullier-Albenque for pointing out the full import of Ref. [\onlinecite{rul09a}]. Work at the Ames Laboratory was supported by the U.S. Department of Energy — Basic Energy Sciences
under Contract No. DE-AC02-07CH11358.
\end{acknowledgments}

\clearpage

\clearpage

\clearpage

\clearpage

\clearpage

\clearpage

\clearpage

\begin{thebibliography}{99}

\bibitem{kam08a} Y. Kamihara, T. Watanabe, M. Hirano, and H. Hosono, J. Am. Chem. Soc. {\bf 130}, 3296 (2008).

\bibitem{che08a} X. H. Chen, T. Wu, G. Wu, R. H. Liu, H. Chen and D. F. Fang, Nature {\bf 453}, 761 (2008).

\bibitem{ren08a} Z.-A. Ren, G.-C. Che, X.-L. Dong, J. Yang, W. Lu, W. Yi, X.-L. Shen, Z.-C. Li, L.-L. Sun, F. Zhou and Z.-X. Zhao, Europhys. Lett. {\bf 83}, 17002 (2008).

\bibitem{rot08a} M. Rotter, M. Tegel, and D. Johrendt, Phys. Rev. Lett. {\bf 101}, 107006 (2008).

\bibitem{sef08a} Athena S. Sefat, Rongying Jin, Michael A. McGuire, Brian C. Sales, David J. Singh, and David Mandrus, Phys. Rev. Lett. {\bf 101}, 117004 (2008).

\bibitem{nin08a} N. Ni, M. E. Tillman, J.-Q. Yan, A. Kracher, S. T. Hannahs, S. L. Bud'ko, and P. C. Canfield, Phys. Rev. B {\bf 78}, 214515 (2008).

\bibitem{chu09a} Jiun-Haw Chu, James G. Analytis, Chris Kucharczyk, and Ian R. Fisher, Phys. Rev. B {\bf 79}, 014506 (2009).

\bibitem{nin09a} F. L. Ning, K. Ahilan, T. Imai, A. S. Sefat, R. Jin, M. A. McGuire, B. C. Sales, D. Mandrus, J. Phys. Soc. Jpn. {\bf 78}, 013711 (2009).

\bibitem{lil09a} L. J. Li, Y. K. Luo, Q. B. Wang, H. Chen, Z. Ren, Q. Tao, Y. K. Li, X. Lin, M. He, Z. W. Zhu, G. H. Cao, and Z. A. Xu, New J. Phys. {\bf 11}, 025008 (2009).

\bibitem{can09a} P. C. Canfield, S. L. Bud'ko, Ni Ni, J. Q. Yan, and A. Kracher,  Phys. Rev. B {\bf 80}, 060501 (2009).

\bibitem{nin09b} N. Ni, A. Thaler, A. Kracher, J. Q. Yan, S. L. Bud'ko, and P. C. Canfield, Phys. Rev. B {\bf 80}, 024511 (2009).

\bibitem{pra09a} D. K. Pratt, W. Tian, A. Kreyssig, J. L. Zarestky, S. Nandi, N. Ni, S. L. Bud'ko, P. C. Canfield, A. I. Goldman, R. J. McQueeney, arXiv:0903.2833v1, unpublished.

\bibitem{les09a} C. Lester, Jiun-Haw Chu, J. G. Analytis, S. C. Capelli, A. S. Erickson, C. L. Condron, M. F. Toney, I. R. Fisher, and S. M. Hayden, Phys. Rev. B {\bf 79}, 144523 (2009).

\bibitem{mun09a} Eun Deok Mun, et al., in preparation.

\bibitem{nin08b} N. Ni, S. L. Bud'ko, A. Kreyssig, S. Nandi, G. E. Rustan, A. I. Goldman, S. Gupta, J. D. Corbett, A. Kracher, and P. C. Canfield, Phys. Rev. B {\bf 78}, 014507 (2008).

\bibitem{tan09a} M. A. Tanatar, N. Ni, C. Martin, R. T. Gordon, H. Kim, V. G. Kogan, G. D. Samolyuk, S. L. Bud'ko, P. C. Canfield, and R. Prozorov, Phys. Rev. B {\bf 79}, 094507 (2009).

\bibitem{tan09b} M. A. Tanatar, A. Kreyssig, S. Nandi, N. Ni, S. L. Bud'ko, P. C. Canfield, A. I. Goldman, and R. Prozorov, Phys. Rev. B {\bf 79}, 180508 (2009).

\bibitem{fan09a} Lei Fang, Huiqian Luo, Peng Cheng, Zhaosheng Wang, Ying Jia, Gang Mu, Bing Shen, I. I. Mazin, Lei Shan, Cong Ren, and Hai-Hu Wen, arXiv:0903.2418v1, unpublished.

\bibitem{rul09a} F. Rullier-Albenque, D. Colson, A. Forget, and H. Alloul, Phys. Rev. Lett. {\bf 103}, 057001 (2009).

\bibitem{var89a} A. A. Varlamov, V. S. Egorov, and A. V. Pantsulaya, Adv. Phys. {\bf 38}, 469 (1989).

\bibitem{liv99a} D. V. Livanov, Phys. Rev. B {\bf 60}, 13439 (1999).

\end{thebibliography}
\end{document}